\documentclass[final,5p,times,twocolun]{elsarticle}
\usepackage{graphicx}
\usepackage{amssymb}
\usepackage[amssymb]{SIunits}
\usepackage{pdfpages}
\usepackage{subfigure}
\usepackage{float}
\usepackage{amsmath}
\usepackage{color}
\usepackage{ulem}
\usepackage{lineno}

\usepackage{xspace}
\usepackage{booktabs}
\usepackage{multirow}

\begin{document}
\begin{frontmatter}
\title{Building a large-area GEM-based readout chamber for the upgrade of the ALICE TPC}
\address[tum]{Physik Department E62, Technische Universit\"{a}t M\"{u}nchen, Garching, Germany}
\address[clu]{Excellence Cluster 'Origin and Structure of the Universe', Garching, Germany}

\author[tum,clu]{P.~Gasik\corref{cor1}}
\author[]{for the ALICE Collaboration}

\cortext[cor1]{Corresponding author: p.gasik@tum.de}

\begin{abstract}
A large Time Projection Chamber (TPC) is the main device for tracking and charged-particle identification in the ALICE experiment at the CERN LHC. After the second long shutdown in 2019-2020, the LHC will deliver Pb beams colliding at an interaction rate up to 50 kHz, which is about a factor of 100 above the present read-out rate of the TPC. To fully exploit the LHC potential the TPC will be upgraded based on the Gas Electron Multiplier (GEM) technology. 

A prototype of an ALICE TPC Outer Read-Out Chamber (OROC) was equipped with twelve large-size GEM foils as amplification stage to demonstrate the feasibility of replacing the current Multi Wire Proportional Chambers with the new technology. With a total area of $\sim$0.76\,m$^2$ it is the largest GEM-based detector built to date.
The GEM OROC was installed within a test field cage and commissioned with radioactive sources. 
\end{abstract}

\begin{keyword}
LHC\sep ALICE TPC\sep GEM TPC\sep Gas Electron Multiplier\sep continuous read-out
%% keywords here, in the form: keyword \sep keyword

%% PACS codes here, in the form: \PACS code \sep code

%% MSC codes here, in the form: \MSC code \sep code
%% or \MSC[2008] code \sep code (2000 is the default)	

\end{keyword}

\end{frontmatter}
%\linenumbers

%%\graphicspath{{../figures/}{../figures/trigger/}{../figures/results/}{../figures/electronics/}

\section{Introduction}
\label{sec:intro}
ALICE (A Large Ion Collider Experiment) at the CERN LHC is planning a major upgrade of the central barrel detectors, including the TPC, to cope with the foreseen increase of the LHC luminosity after 2020. The current Multi Wire Proportional Chambers \cite{bib:alicetpc} will be replaced by the continuously operated detectors employing Gas Electron Multiplier (GEM) technology \cite{bib:sauli}. The new readout chambers are designed to read all minimum bias Pb-Pb events that the LHC will deliver at the anticipated peak interaction rate of 50 kHz. This will result in a significant improvement on the sensitivity to rare probes that are considered key observables to characterise the QCD matter created in heavy-ion collisions.

Ungated operation of the ALICE TPC will lead to a considerable accumulation of positive ions in the drift volume that emerge from the amplification region. The resulting space-charge distortions must be kept within limits that allow
efficient online track reconstruction and distortion corrections. To fulfil the challenging requirements of the upcoming upgrade, a novel configuration of GEM detectors has been developed. It allows us to maintain excellent particle identification and efficient ion trapping by stacking four GEM foils operated at a specific field configuration. 

To achieve the required Signal-to-Noise ratio the GEM stacks will be operated at an effective gain of 2000. In order to keep the resulting drift field distortions at this gain within a tolerable level, an upper limit for the ion backflow of
1\% has been set. At the same time, the local energy resolution of the read-out chambers must not exceed $\sigma$/E = 12\% at the 5.9\,keV peak of $^{55}$Fe in order to retain the performance of the existing system in terms of d$E$/d$x$ resolution. The upgraded TPC will be operated with a Ne-CO$_2$-N$_2$ (90-10-5) gas mixture.
%\subsection{Quadruple GEM configuration}

The use of Gas Electron Multiplier foils offers an intrinsic ion backflow ($IB$) suppression. $IB$ is defined as a ratio of the (ion) current measured at the cathode divided by the (electron) current measured at the readout anode. Extensive tests were carried out with different kinds of small prototypes to measure the $IB$ as a function of different voltage settings and gas mixtures. The optimal performance can be obtained employing a four-GEM stack using foils with different optical transparencies, i.e. varying the pitch between GEM holes \cite{bib:tdr}. A performance within specifications has been achieved with a quadruple GEM system in which the foils layers 1 and 4 have a standard hole pitch (S, 140\,$\mu$m), whereas the foils in layers 2 and 3 have a hole pitch that is two times larger (LP, 280\,$\mu$m). In order to keep the $IB$ uniform across the detector area, the foils are rotated with respect to each other to avoid unintended alignment of the holes between the foils \cite{bib:tdradd}.
\begin{figure}[!htb]
\begin{center}
		\includegraphics[width=0.9\linewidth]{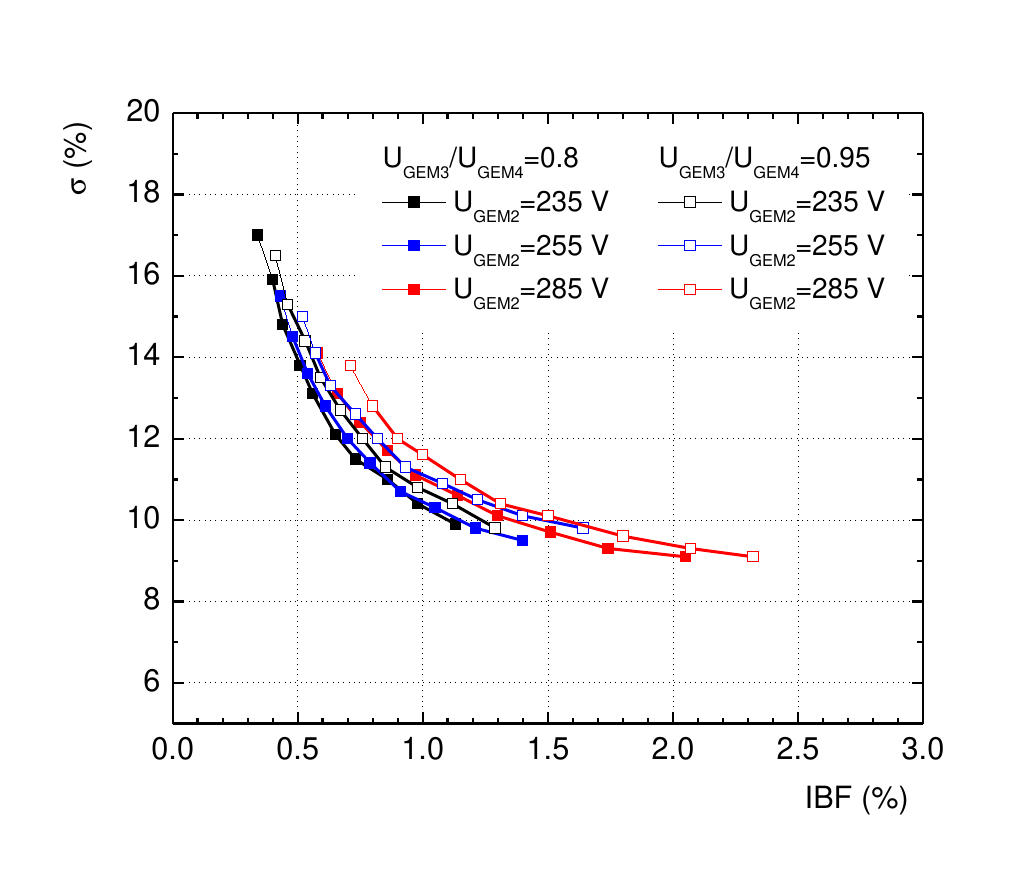}
		\caption{Energy resolution at 5.9 keV and ion backflow for different GEM stack voltages \cite{bib:tdr}.}
		\label{fig:ibf}
		\end{center}
\end{figure}
The $IB$ in the S-LP-LP-S system can be reduced to the level of $\sim$0.65\% keeping the energy resolution at the acceptable level of $\sim$12\% as measured with an $^{55}$Fe source (see fig.\ref{fig:ibf}).

In addition, extensive discharge probability studies have been performed to assure the stability of the chosen solution. Upper limits for the discharge probability upon irradiation with alpha particles (at a gas gain of 2000) are found to be of the order of 10$^{-10}$ per alpha particle. This number is compatible with that for standard triple GEM systems that are operated routinely in high-rate experiments \cite{bib:tdradd}.

In the course of the R\&D program, several full-size prototypes  of the Inner Readout Chamber (IROC) equipped with GEM foils have been built \cite{bib:tdr, bib:tdradd,  bib:3gemiroc, bib:vci13}. The d$E$/d$x$ resolution of the four-GEM IROC was evaluated at the test-beam at CERN/PS with a beam of 1 GeV/$c$ electrons and pions. The relative energy resolution was measured to be $<10\%$, comparable with the currently operated MWPC-based chambers. In a test-beam at the CERN/SPS, the detector was exposed to an intense hadron flux from a secondary production target. The measured discharge probability of $(6\pm4)\times10^{-12}$ per incoming hadron is compatible with an efficient and safe operation of the TPC in RUN 3 and beyond \cite{bib:tdradd}.
\section{Full-size Outer Read-Out Chamber prototype}
In order to check the feasibility of the upgrade solution, a full-scale prototype of an Outer Read-Out Chamber (OROC) was assembled and successfully operated in 2015, being the largest detector of this type to date. The main goal was to establish the stages of GEM integration on a large-size chamber body.

The chamber is of trapezoidal shape of 114\,cm length and 47\,cm (87\,cm) width of the short (long) parallel sides. Due to technical reasons, it is not possible to produce GEM foils of the size of the OROC chamber. Therefore, three independent GEM stacks are mounted to fit the required dimensions. 

\begin{figure}
\centering
\includegraphics[width=0.9\linewidth]{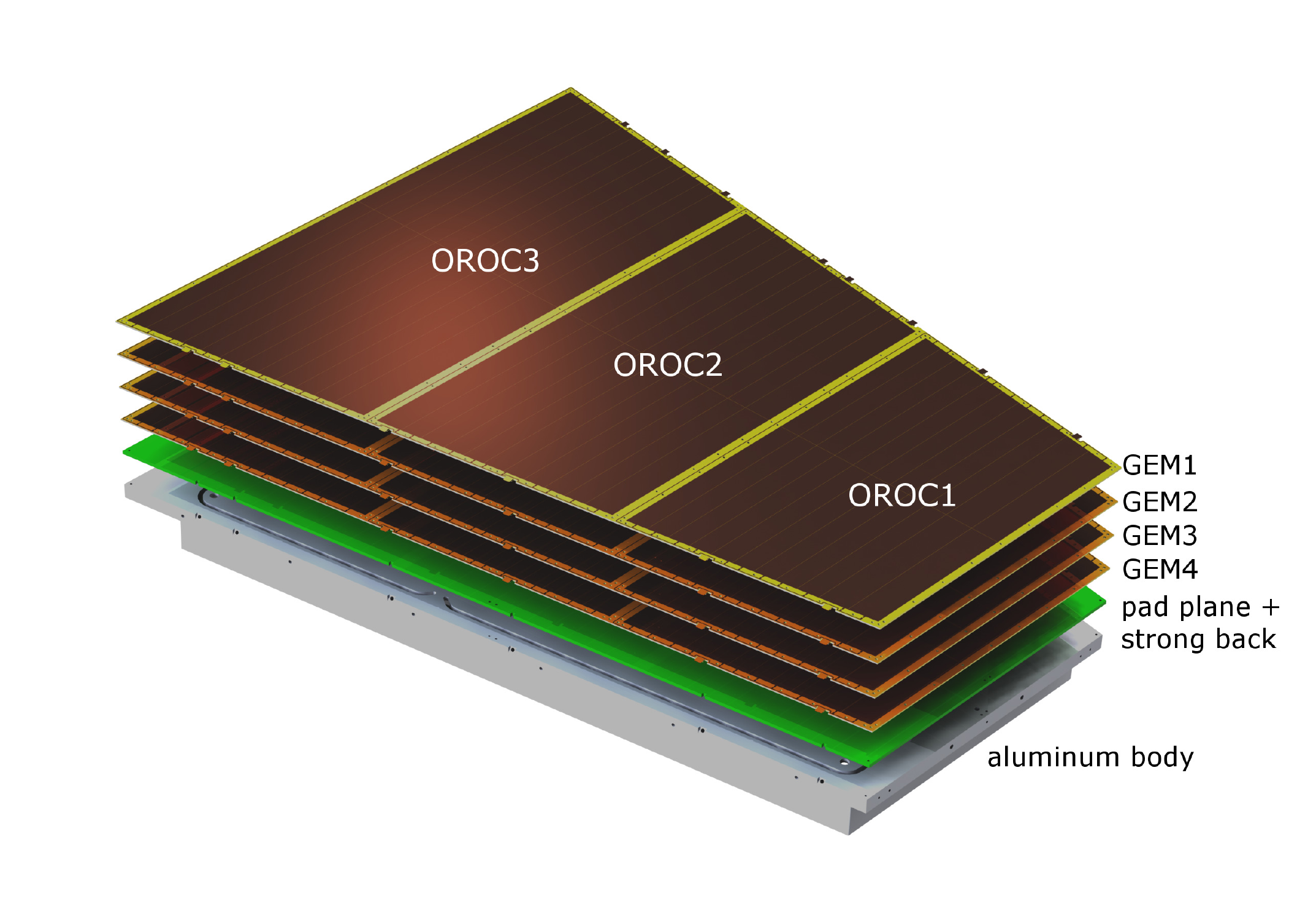}
\caption[Exploded view of the OROC prototype]{Exploded view of the OROC prototype and its main components.}
\label{fig:oroc}
\end{figure}

The prototype is assembled on a spare OROC of the ALICE TPC \cite{bib:alicetpc}. The mechanical structure of the chamber consists of four main components: the GEM stacks, the pad plane ($\sim10000$ pads of rectangular 6$\times$10 and 6$\times$15\,mm$^2$ shape), made of a multi-layer Printed Circuit Board (PCB), an additional $3\,\mathrm{mm}$ Stesalit insulation plate (so called 'strong back') and an aluminum frame (alubody). A schematic view of a prototype together with its main components is shown in fig.~\ref{fig:oroc}. 

The following sections present a detailed description of the GEM foil design, quality control and the detector assembly procedure. It was initially based on the experience gained with the prototypes of the Inner Readout Chamber, however, many modification have been implemented to improve the design and the assembly methods.

\subsection{GEM foils}
Twelve large-area GEM foils have been produced at CERN using the single-mask technique \cite{bib:singlemask}. They have trapezoidal shape, as depicted in fig.~\ref{fig:oroc}. The dimensions of the OROC GEM foils, numbered according to their size OROC1,2,3 (see fig.~\ref{fig:oroc}), are summarised in table~\ref{tab:orocsizes}.
\begin{table}[hbt]\footnotesize
    \caption{Size of the OROC foils used in the prototype. }
  \centering
  \begin{tabular}{lccc}
    \hline
       GEM 	type		& size 					& no. HV segments 	& av. segment size 	\\ 
    				& H (cm) $\times$ W (cm)		&				& (cm$^2$)			\\\hline
    OROC1			& 37.1 $\times$ 59.7 (46.8)	& 	22			& 120	\\
    OROC2			&  37.2 $\times$ 73.0 (59.7)	&	22			& 100	\\
    OROC3		 	&  39.9 $\times$ 87.0 (73.0)	&	24			& 80		\\ \hline
  \end{tabular}
  \label{tab:orocsizes}
\end{table}
In order to reduce the total charge stored in the GEM foil, one side of the foil is segmented into sectors with separate HV supply with a surface area of approximately 100\,cm$^2$. The number of HV segments and their average area is given in table~\ref{tab:orocsizes}. The gap between HV segments is $200\,\mu\mathrm{m}$. An additional $100\,\mu\mathrm{m}$ of copper between the sector boundary and the GEM holes is added on each side of the gap to account for possible misalignments during the production of the foil. 
Following the baseline configuration of the detector (see sec. \ref{sec:intro}) each GEM foil in a stack differs in the pitch between the holes and in the rotation of the GEM hole pattern, according to table~\ref{tab:foildet}.
\begin{table}[hbt]\footnotesize
    \caption{Specification of GEM foils in a quadruple stack of the prototype}
  \centering
  \begin{tabular}{ccc}
    \hline
       Foil in a stack	& pitch ($\mu$m)					& rotation \\\hline
	GEM1		& 140 & 0$^{\circ}$ \\
	GEM2			& 280 & 90$^{\circ}$ \\
	GEM3			& 280 & 0$^{\circ}$ \\
	GEM4 	& 140 & 90$^{\circ}$ \\\hline
  \end{tabular}
  \label{tab:foildet}
\end{table}
The HV distribution trace runs along three sides of each foil. It consists of a 1 mm copper path and connection flaps. Sectors are powered in parallel via loading resistors soldered directly on the foil between the distribution path and a sector (see sec.~\ref{sec:hv}). The bottom (non-segmented) side of the foil is connected directly to the HV, therefore no distribution path is needed.

The GEM foils are tested before and after framing in order to validate them for the final assembly. Quality assurance (QA) tests performed with the prototype foils include  optical and high voltage checks described below.
\subsubsection{Optical check}
\label{sec:qa}
The optical test allows to check the hole size uniformity across the GEM active area and to record any mechanical defects on the foils in order to identify possible correlations with electrical instabilities which could appear during  detector operation. Large defects on the copper or polyimide layers may increase the probability of electrical breakdown. Mechanical defects, such as cuts, may result in short circuits between top and bottom side of a foil.

The overall quality of the GEM foils used with the prototype is fairly good. No severe defects were found. All small defects found on the foils were photographed for the report. 
\subsubsection{HV test}
A HV test is performed at each step of the detector assembly. The aim of the HV test is to check if the foils are stable under high voltage, measure leakage currents of sectors, identify segments with short circuits and find correlations between stability under HV and defects observed in the foil sectors.

All foils are tested in air before and after framing. Each segment is ramped up directly to the test potential of $U_{\mathrm{QA}}=600$\,V with a high current limit of $I_{\mathrm{limit, QA}}=4\,\mu\mathrm{A}$, to let the high current burn or evaporate foil contaminants (dust, residuals of chemicals used in the production process). In parallel, the discharge rate and the approximate position of each spark is recorded. 

In case of a high leakage current or high-rate discharging observed at the same position of a HV segment, the HV test is considered as failed and the foil is sent back to CERN for reprocessing (usually additional cleaning) or, in case of fatal defect, not accepted for the further assembly stages. In total, $\sim$15\% foils (2 out of 12 received) did not pass the HV test and were returned to the producer. After the additional cleaning procedure their performance is acceptable.

Before assembling the detector, all sectors in all foils were stable and had low and accep\-ta\-ble leakage currents (i.e. below 0.5\,nA).
\subsection{Supporting frames}
\label{sec:frames}
The foils are glued on $2\,\mathrm{mm}$ thick Permaglas frames and then mounted in a stack. The frames contain a $400\,\mu\mathrm{m}$ thick spacer grid to prevent the foils to approach each other due to electrostatic forces. The grid pattern follows the segmentation of the GEM foil and is aligned with the HV segment boundaries. To evaluate the necessity of the spacer grid, it was removed from the OROC3 frames.

When the stack is mounted, the loading resistors fit into grooves milled in the bottom side of the frame placed above.
Framed foils, therefore, lie flat on top of each other.
\subsection{Detector assembly}
\label{sec:alubody}
\subsubsection{Gluing the foils}
Before gluing onto the supporting frames, the foils are stre\-tched using a modified commercial, pneumatic stretching tool with a tension of 10\,N/cm. Once the foil is stretched, it is positioned on its frame and aligned with metal pins. A heavy aluminum plate (milled in a way to prevent it from touching the active area  of the foil) is used to press the foil onto the frame. The epoxy glue used is ARALDITE 2011\texttrademark. After gluing, the raw material surrounding the GEM foil is cut off and the framed foil is tested under high voltage.
\subsubsection{Mounting the GEM stack}
After the HV check of the framed foils, the loading resistors (SMD 1206) are soldered and the foils are mounted in three stacks on the alubody of the chamber (unsectorized side facing the pad plane). The foils in a stack are screwed to the alubody with nylon screws reinforced with 25\% glass fibres, thus each foil can be easily exchanged at any time if needed. On the corners, alignment pins are used to keep all the frames in the same position with a precision of $<100\,\mu\mathrm{m}$. The distance between subsequent foils is 2\,mm, given by the thickness of the supporting frames. HV is applied to the foils via Kapton-coated wires soldered to the HV flaps on each foil.  
\subsubsection{Test box with a field cage}
The chamber is mounted in a test box which contains a drift cathode and a rectangular field cage with outer dimensions of $92.4\,\textrm{cm}\times119.4\,\mathrm{cm}$.
The drift electrode is made of $50\,\mu\mathrm{m}$ aluminised Kapton foil. The field cage has 8 field-defining strips with a pitch of 15\,mm. A $1\,\mathrm{M}\Omega$ resistor is connected between each of the strips (0.5 M$\Omega$ between the drift electrode and a first strip, due to the twice shorter distance). 

The drift distance to the first GEM is 11.1\,cm. The last strip of the field cage is grounded via a resistor adjusted to match its potential with a drift field at the top electrode of GEM1. 

\subsection{High Voltage supply}
\label{sec:hv}
Each GEM stack of the prototype is powered using an independent, external resistor chain supplied by a channel of the HV power supply. Resistor values are given by the so-called "baseline settings" optimised during the R\&D period for a low ion backflow, good energy resolution and low discharge probability \cite{bib:tdradd}. The settings can be scaled (using an overall Scaling Factor, $SF$) in order to vary the total gain. $SF=100\%$ corresponds to the nominal gain of 2000 in Ne-CO$_2$-N$_2$ (90-10-5) gas mixture. All together, 5 HV channels are needed to power up the prototype (3 channels for GEM stacks, one for the drift electrode and one for the last strip of the field cage).

Protection resistors ($R_{\mathrm{p}}$) are installed on the top side of each foil. This assures that in case of a discharge across a foil, the voltage drop occurs only on the top side, whereas the bottom side stays at its nominal potential. This helps to prevent the propagation of the discharge to the next foil or to the pad plane and readout electronics. 
Moreover, in case of occasional sparks, the protection resistors limit the current supplied by the power supply and quench the discharges. They also decrease the current flowing through the sector in case of a short circuit between top and bottom sides of the foil.
The values of $R_{\mathrm{p}}$ were chosen keeping in mind the current density expected in the future Pb-Pb collisions at 50 kHz at a gain of 2000, which is roughly $10\,\mathrm{nA}/\mathrm{cm}^2$. Such a current may result in a significant potential drop across a large loading resistor, thus reducing the gain. Therefore $10\,\mathrm{M}\Omega$ resistors were chosen for GEM1-3 and  $1\,\mathrm{M}\Omega$ for GEM4, where the expected amplification current is the largest
\section{Commissioning with $^{55}$Fe source}
\label{sec:comm}
The prototype in the test box was tested in the lab with an Ar-CO$_2$ (82-18) gas mixture and irradiated with a $^{55}$Fe X-ray source. For the commissioning of the detector, 256 pads ($\sim$75\,cm$^2$) were connected to a charge-sensitive preamplifier and the signal was digitised with a multi-channel analyser for measuring the X-ray spectrum. A picoamperemeter is used to estimate the gain by measuring the current in the pad plane taking into account the rate of X-rays. 

\begin{figure}[hbt]
\centering
\includegraphics[width=0.87\linewidth]{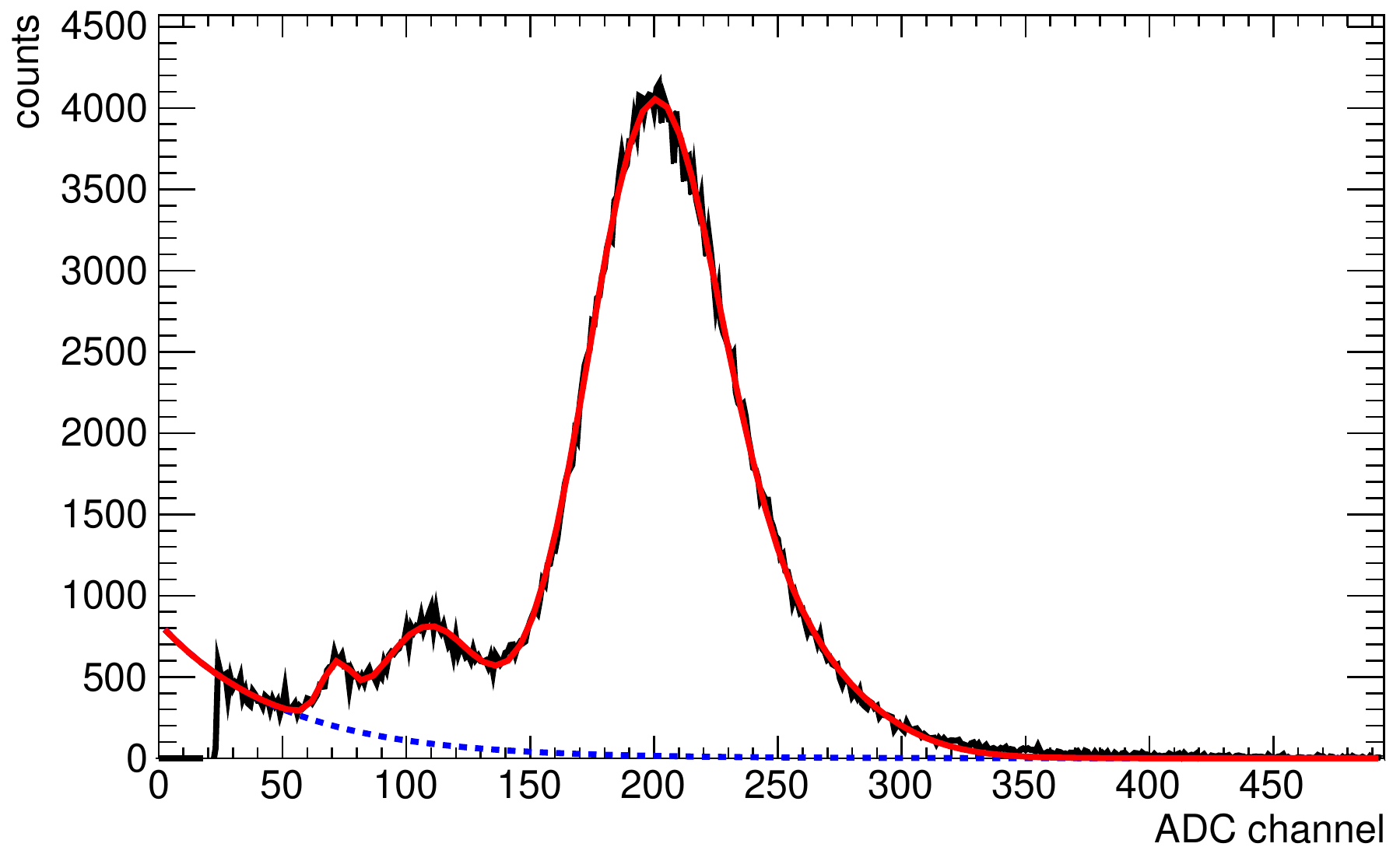}
\caption[$^{55}$Fe spectrum obtained in Ar-CO$_2$ (82-18)]{$^{55}$Fe spectrum obtained in Ar-CO$_2$ (82-18) with OROC2 stack.}
\label{fig:fe55}
\end{figure}

Figure~\ref{fig:fe55} shows an example of the energy spectrum of $^{55}$Fe obtained at 128\% of the baseline HV settings, which correspond to a gain of around 80000. Both argon peaks are visible together with a low intensity peak, centred on ADC channel 70, corresponding to the aluminum fluorescence X-rays (source was placed in front of the aluminised drift electrode). The energy resolution of the main peak is $\sigma$/E($^{55}$Fe)\,=\,11.8\%. It should be noted that the drift field used in this studies was set to $\sim$180\, V/cm, which is a factor more than twice lower than the nominal ALICE drift field (400\,V/cm).
\section{Mechanical studies of the spacer frame}
The introduction of a spacer grid to the frames is justified by considerations of the electrostatic forces between adjacent foils. As it increases the complexity of the assembly and introduces detection inefficiencies, the spacer grid of all OROC3 frames was removed in order to prove its necessity. In fact, during commissioning, frequent discharges were observed in this stack above $\sim$80\% of the nominal high voltage settings. As this is a clear hint for discharges due to the sag between adjacent foils, a sagging test was conducted with only GEM4 mounted on the pad plane.

The induction field was successively increased, while keeping the potential across the GEM at zero. Above an electric field of $\sim$ 3\,kV/cm, the foil sag towards the grounded pad plane was visible by eye and frequent discharges could be observed. Quantitative measurements based on the observation of the movement of the reflection of a laser pointed to the middle of the GEM foil were performed. 

In order to define the number of ribs in a spacer grid, at first only one transverse bar was placed between the pad plane and the GEM foil. While the foil sag could still be observed, discharges towards the pad plane were already prevented at nominal voltage. With a simple cross (one longitudinal and one transverse bar) placed onto the pad plane, dividing the OROC3 area into quarters, $\sim$800\,cm$^2$ each, the sag was unperceivable and it was possible to operate the detector stably up to the nominal voltage with a vast safety margin. The outcome of these studies has been included in the final modifications of the readout chamber design.

\section{Summary and Outlook}
A GEM-based prototype of an Outer Readout Chamber for the ALICE TPC Upgrade was built and commissioned in order to check the feasibility of the proposed technology to be used on a large scale. Three quadruple GEM stacks were mounted as an amplification stage instead of the currently used wire planes. The design of the readout chambers has been validated and the assembly procedures were exercised and improved for the final production.

The start of the production of the readout chambers for the upgraded TPC is planned for middle of 2016. All  readout chambers are planned to be assembled until 2019, when the LS2 is supposed to start and the TPC will be moved to the surface. Installation and commissioning of the chambers will last until the end of 2020.
\section*{Acknowledgements}
This research was supported by the DFG cluster of excellence 'Origin and Structure of the Universe' (www.universe-cluster.de). 
The author would like to thank the participants of the "School of ROC" workshop, in which the OROC prototype was built.

\end{document}